\pdfoutput=1
\documentclass[11pt,twoside,a4paper,cmspaper,final,collab]{cms-tdr}
\def\svnVersion{31e82c8}\def\svnDate{2024/09/04}\def\cmsCernNoTag{CERN-EP-2022-280}\def\cmsCernDate{\today}\def\cmsMessage{Published in Physics Letters B as \href{http://dx.doi.org/10.1016/j.physletb.2024.138936}{\doi{10.1016/j.physletb.2024.138936}.}}
\begin{document}\cmsNoteHeader{HIN-21-006}

\newcommand{\PbPb}{\ensuremath{\text{PbPb}}\xspace}
\newcommand{\AuAu}{\ensuremath{\text{AuAu}}\xspace}
\newcommand{\pp}{\ensuremath{\Pp\Pp}\xspace}
\newcommand{\pPb}{\ensuremath{\Pp\text{Pb}}\xspace}
\newcommand{\KS}{\PKzS}
\newcommand{\PKZ}{\PKz}
\newcommand{\PKAZ}{\PAKz}
\newcommand{\LM}{\PGL}
\newcommand{\ALM}{\PAGL}
\newcommand{\KSKS}{\ensuremath{\PKzS\PKzS}\xspace}
\newcommand{\LLALAL}{\ensuremath{\PGL\PGL}\xspace}
\newcommand{\ALMALM}{\ensuremath{\PAGL\PAGL}\xspace}
\newcommand{\LKS}{\ensuremath{\PGL\PKzS}\xspace}
\newcommand{\ALKS}{\ensuremath{\PAGL\PKzS}\xspace}
\newcommand{\LALKS}{\ensuremath{\PGL\PKzS}\xspace}
\newcommand{\VZ}{\ensuremath{\textrm{V}^{0}}\xspace}
\newcommand{\roots}{\ensuremath{\sqrt{s}}\xspace}
\newcommand{\minv}{\ensuremath{m_{\text{inv}}}\xspace}
\newcommand{\qinv}{\ensuremath{q_{\text{inv}}}\xspace}
\newcommand{\Rin}{\ensuremath{R_{\text{inv}}}\xspace}
\newcommand{\cobs}{\ensuremath{C^{\text{obs}}(\qinv)}\xspace}
\newcommand{\css}{\ensuremath{C(\qinv)}\xspace}
\newcommand{\aobs}{\ensuremath{A^{\text{obs}}(\qinv)}\xspace}
\newcommand{\bobs}{\ensuremath{B^{\text{obs}}(\qinv)}\xspace}
\newcommand{\ass}{\ensuremath{A^{\text{ss}}(\qinv)}\xspace}
\newcommand{\bss}{\ensuremath{B^{\text{ss}}(\qinv)}\xspace}
\newcommand{\asb}{\ensuremath{A^{\text{sb}}(\qinv)}\xspace}
\newcommand{\abb}{\ensuremath{A^{\text{bb}}(\qinv)}\xspace}
\newcommand{\fss}{\ensuremath{f^{\text{ss}}}\xspace}
\newcommand{\fsb}{\ensuremath{f^{\text{sb}}}\xspace}
\newcommand{\fbb}{\ensuremath{f^{\text{bb}}}\xspace}
\newcommand{\cqs}{\ensuremath{C_{\text{QS}}(\qinv)}\xspace}
\newcommand{\cfsi}{\ensuremath{C_{\text{FSI}}(\qinv)}\xspace}
\newcommand{\He}{\ensuremath{^{\;\;\;\;6}_{\LLALAL}{\mathrm{He}}}\xspace}
\newcommand\ket[1]{\mathinner{\vert{#1}\rangle}}
\newlength\cmsTabSkip\setlength{\cmsTabSkip}{1ex}

\cmsNoteHeader{HIN-21-006}

\title{\texorpdfstring{\PKzS}{PKzS} and \texorpdfstring{\LM(\ALM)}{\LM(\ALM)} two-particle femtoscopic correlations in \PbPb collisions at \texorpdfstring{$\sqrtsNN = 5.02\TeV$}{sqrtsNN = 5.02 TeV}}

\author*[cern]{The CMS Collaboration}

\date{\today}

\abstract{ 
Two-particle correlations are presented for \KS, \LM, and \ALM strange hadrons as a function of relative momentum in lead-lead collisions at a nucleon-nucleon center-of-mass energy of 5.02\TeV. The dataset corresponds to an integrated luminosity of 0.607\nbinv and was collected using the CMS detector at the CERN LHC. These correlations are sensitive to quantum statistics and to final-state interactions between the particles. The source size extracted from the \KSKS correlations is found to decrease from 4.6 to 1.6\unit{fm} in going from central to peripheral collisions. Strong interaction scattering parameters (\ie, scattering length and effective range) are determined from the \LALKS and \LLALAL (including their charge conjugates) correlations using the Lednick{\'y}--Lyuboshitz model and are compared to theoretical and other experimental results.
}

\hypersetup{
pdfauthor={CMS Collaboration},
pdftitle={K and Lambda two-particle femtoscopic correlations in PbPb collisions at sqrt(sNN) = 5.02 TeV},
pdfsubject={CMS},
pdfkeywords={CMS, heavy ion, V0, femtoscopy, scattering parameters}} 

\maketitle

\section{Introduction}
\label{sec:intro}
Two-particle correlations in relative momentum, so-called femtoscopic correlations, arising from relativistic heavy ion collisions provide a powerful tool for studying the spatiotemporal characteristics of the particle emitting sources created in these collisions and the subsequent interactions of the emitted particles~\cite{Lisa:2005dd}. All two-particle correlations involving hadrons are affected by final-state interaction (FSI) effects, and correlations of identical particles are also sensitive to the constraints of quantum statistics (QS). The correlations among the neutral \KS, \LM, and \ALM particles, collectively referred to as \VZ particles, are of special interest. Because of their relatively large mass, femtoscopy based on \KS particles is less affected by resonance decays than is the case with more commonly studied pion and charged kaon pairs~\cite{PhysRevC.97.064912}. The results from \KSKS, \LALKS, and \LLALAL correlation studies can be used to determine the strong-interaction scattering parameters, \ie, the scattering length and the effective range, that are impossible to obtain from currently achievable scattering experiments~\cite{PhysRevC.48.792, DESWART1962458, ENGELMANN1966587, EISELE1971204, PhysRev.175.1735}. This information can be used to constrain hadron-hadron correlation models that are used, for example, in modeling the composition of neutron stars~\cite{KSNeutron,PhysRevLett.89.171101}.

Regarding the scattering parameters, of particular interest is establishing whether the interaction between two \LM particles allows for the existence of the H-dibaryon, a bound state with quantum numbers $I = 0$, $J^P = 0^+$, and $S = -2$. In 1977, R. L. Jaffe predicted the existence of such a six-quark ($\PQu\PQu\PQd\PQd\PQs\PQs$) state having a mass about 81\MeV below the threshold of twice the \LM mass by considering the strong attraction resulting from color magnetic interactions~\cite{PhysRevLett.38.195}. Although a double hypernucleus, \He, was subsequently observed in the NAGARA event from the E313 hybrid emulsion experiment at KEK~\cite{PhysRevLett.87.212502,10.1143/PTPS.185.335}, the observed \LLALAL binding energy was not consistent with the conjectured H-dibaryon~\cite{PhysRevLett.110.222002}. A study of \LLALAL correlations may provide additional information on whether the baryon-baryon interaction can lead to the formation of the conjectured H-dibaryon.

Recently, the ALICE Collaboration reported on $\LM\PK$ correlations in lead-lead (\PbPb) collisions at a center-of-mass energy per nucleon pair of $\sqrtsNN = 2.76\TeV$~\cite{PhysRevC.103.055201}. According to their findings, the strong force is repulsive in $\LM\PKp$ interactions, yet attractive in $\LM\PKm$ interactions. For the \LKS pairs, the uncertainty of the ALICE results does not permit a definite conclusion on whether the associated strong interaction is repulsive or attractive. A more precise measurement of \LKS correlations should improve our understanding of the strong interaction in baryon-meson systems.  

This Letter presents \KSKS, \LALKS, and \LLALAL femtoscopic correlations as a function of relative momentum in \PbPb collisions at $\sqrtsNN = 5.02\TeV$, using data recorded by the CMS experiment during the 2018 LHC run. The \KSKS, \LALKS, and \LLALAL correlations are measured in an integrated centrality range 0--80\%, where centrality refers to the percentage of the total inelastic hadronic nucleus-nucleus cross section~\cite{Loizides:2017ack}, and 0\% corresponds to the maximum overlap of the colliding nuclei. This is the first measurement of the \LLALAL femtoscopic correlation in \PbPb collisions at the LHC\@. In addition, the \KSKS correlations are measured in six centrality intervals in the range of 0--60\%. The source size and strong interaction parameters are determined using the Lednick{\'y}--Lyuboshitz (LL) model~\cite{Lednicky:1981su}. Unless otherwise indicated, all measurements include the charge conjugate states, so \LKS and \LLALAL include \ALKS and \ALMALM, respectively. Tabulated results are provided in the HEPData record for this analysis~\cite{hepdata}.

\section{Experimental setup and data sample}
\label{sec:exp_data}
The central feature of the CMS apparatus is a superconducting solenoid of 6\unit{m} internal diameter, providing a magnetic field of 3.8\unit{T}. Within the solenoid volume there is a silicon pixel and strip tracker, a lead tungstate crystal electromagnetic calorimeter, and a brass and scintillator hadron calorimeter, each composed of a barrel and two endcap sections. The silicon pixel detector~\cite{Phase1Pixel} is composed of 1856 silicon pixel modules distributed in four 54 cm long barrel layers at radii of 2.9--16.0\unit{cm} plus three pairs of endcap disks covering radii of 4.5--16.1\unit{cm} at longitudinal distances of 31--51\unit{cm} from the origin.  The 15\,148 silicon strip modules are arranged in 10 barrel layers at radii of 20--116\unit{cm} plus 3 pairs of small and 9 pairs of large endcap disk layers.  Charged particles of pseudorapidity $\abs{\eta} < 3$ are reconstructed with the combined system. For particles with transverse momentum of $1 < \pt < 10\GeV$, the track resolutions are typically 1.5\% in \pt and 20--75$\mum$ in the transverse impact parameter~\cite{DP-2020-049}. The barrel and endcap detectors are extended to the forward region with two calorimeters which use steel as the absorber and quartz fibers as the sensitive material. These hadron forward (HF) calorimeters are located 11.2\unit{m} from the interaction region, one on each side, and provide coverage in the range $3.0 < \abs{\eta} < 5.2$. These detectors are segmented into multiple $0.175{\times}0.175$ ($\Delta\eta{\times}\Delta\phi$) ``towers'', where $\phi$ is azimuthal angle in radians. Muons are measured in gas-ionization detectors embedded in the steel flux-return yoke outside the solenoid. Events of interest are selected using a two-tiered trigger system. The first level, composed of custom hardware processors, uses information from the calorimeters and muon detectors to select events at a rate of around 100\unit{kHz} within a fixed latency of about 4\mus~\cite{CMS:2020cmk}. The second level, known as the high-level trigger, consists of a farm of processors running a version of the full event reconstruction software optimized for fast processing, and reduces the event rate to around 1\unit{kHz} before data storage~\cite{Khachatryan:2016bia}. A more detailed description of the CMS detector, together with a definition of the coordinate system used and the relevant kinematic variables, can be found in Ref.~\cite{CMS:2008xjf}.

With an integrated luminosity of 0.607\nbinv~\cite{CMS:LUM-17-003, CMS:LUM-18-001}, this analysis uses $4.27\times 10^9$ minimum bias events that are triggered by requiring signals above the readout threshold of 3\GeV in each of the HF calorimeters~\cite{Khachatryan:2016bia}. Background events due to beam-gas interactions and nonhadronic collisions are filtered offline by applying the procedure described in Ref.~\cite{Khachatryan:2016odn}. The events used in this analysis are required to have at least one primary interaction vertex determined using two or more tracks~\cite{Collaboration_2014} within a distance of 15\unit{cm} from the center of the nominal interaction point along the beam axis and to have at least two calorimeter towers in each HF detector with energy deposits of more than 4\GeV per tower. The shapes of the clusters in the pixel detector are required to be compatible with those expected in \PbPb collisions in order to suppress the contamination from events with multiple collisions~\cite{JHEP022010}. The combined trigger and offline selection efficiency for inelastic events is greater than 95\%. The event centrality is obtained from the transverse energy deposited in both HF calorimeters, using the methodology described in Ref.~\cite{Chatrchyan:2011sx}. The analysis makes use of a minimum bias Monte Carlo \PbPb sample, based on the \HYDJET 1.9~\cite{Gale:2013da} event generator with a full detector simulation using \GEANTfour~\cite{GEANT4}.

\section{Reconstruction of \texorpdfstring{\KS}{} and \texorpdfstring{\LM}{} candidates}
\label{sec:rec_v0}
The \KS and \LM candidates, denoted as \VZ candidates, used in this study are reconstructed as in previous CMS analyses~\cite{CMS:2021fhf, Khachatryan:2011tm, Khachatryan:2014jra}. The \VZ candidates are found by combining oppositely charged tracks that pass criteria based on the ``loose'' selection discussed in Ref.~\cite{Collaboration_2014}. The charged tracks are assumed to be $\PGpp\PGpm$ in \KS reconstruction and $\PGpm\Pp$ in \LM reconstruction. For the latter, the higher momentum track is assumed to be a proton since the proton carries nearly all of the momentum in the \LM decay. Each of the oppositely charged tracks must have hits in at least three layers of the silicon tracker, and both tracks must have transverse and longitudinal impact parameter significances (defined as the parameter value divided by its uncertainty) with respect to the primary vertex greater than 1. The two tracks are fitted to a common vertex and the $\chi^2$ per degree of freedom (dof) from the fit must be less than 7. The distance of closest approach between the two tracks is required to be less than 1\unit{cm}. As a consequence of  the long lifetime of \KS and \LM particles, the significance of the \VZ decay length, which is the three-dimensional distance between the primary and \VZ vertices divided by its uncertainty, is required to be greater than 2.5 to reduce combinatorial background contributions. To remove \KS candidates misidentified as \LM particles and vice versa, the \LM(\KS) candidates must have a corresponding $\PGpp\PGpp\,(\Pp\PGpm)$ mass more than 14 (7)\MeV (corresponding to approximately 3 times the average resolution) away from the world-average value~\cite{PDG2020} of the \KS(\LM) mass. The angle $\theta$ between the \VZ momentum vector and the vector connecting the primary and \VZ vertices is required to satisfy $\cos\theta > 0.999$. This reduces the contribution from nuclear interactions, random combinations of tracks, and \LM particles originating from weak decays of \PgX and \PgO particles. 

Further selection of \VZ candidates is performed with a boosted decision tree (BDT)~\cite{Hocker:2007ht}. The selection is optimized separately for \KS and \LM candidates. The discriminating variables include: the collision centrality, the \VZ candidate \pt and rapidity ($y$), the distance of closest approach of the track pair, the three-dimensional decay length and significance, $\cos\theta$, and the \VZ vertex fit $\chi^2$. The included variables related to the \VZ daughters are \pt, uncertainty in \pt, $\eta$, the number of hits in the silicon tracker, the number of pixel detector layers with hits, and the transverse and longitudinal impact parameter significances with respect to the primary vertex. The BDT training is performed using the simulated minimum bias sample separated into the signal and background subsamples using the generator-level information. The \KS mesons are selected with $1 < \pt < 8.5\GeV$ and $\abs{y} < 1$, while the \LM baryons are required to have $1.8 < \pt < 8.5\GeV$ and $\abs{y} < 1$. The minimum \pt and maximum $y$ requirements are used to reduce background while the maximum \pt requirement is to reduce contributions from jets. The combined \VZ reconstruction and selection efficiencies are strongly dependent on the centrality of the event and the \pt of the \VZ. Integrating over the selected \pt ranges, the combined efficiencies from the most central to peripheral \PbPb collisions are 1--3\% for \KS and 1--2\% for \LM. The \VZ reconstruction algorithm does not prevent a track from being used for more than one \VZ. While this is normally an infrequent occurrence, selecting pairs of \VZ particles close together in phase space makes it a significant contribution. To resolve this problem, for each correlation measurement, a check of each pair of \VZ candidates is performed and if two \VZ candidates are found to share one or both daughter tracks, one of the \VZ candidates is randomly selected to be removed from the event.

\begin{figure*}[htbp]
 \centering
 \includegraphics[width=0.49\textwidth]{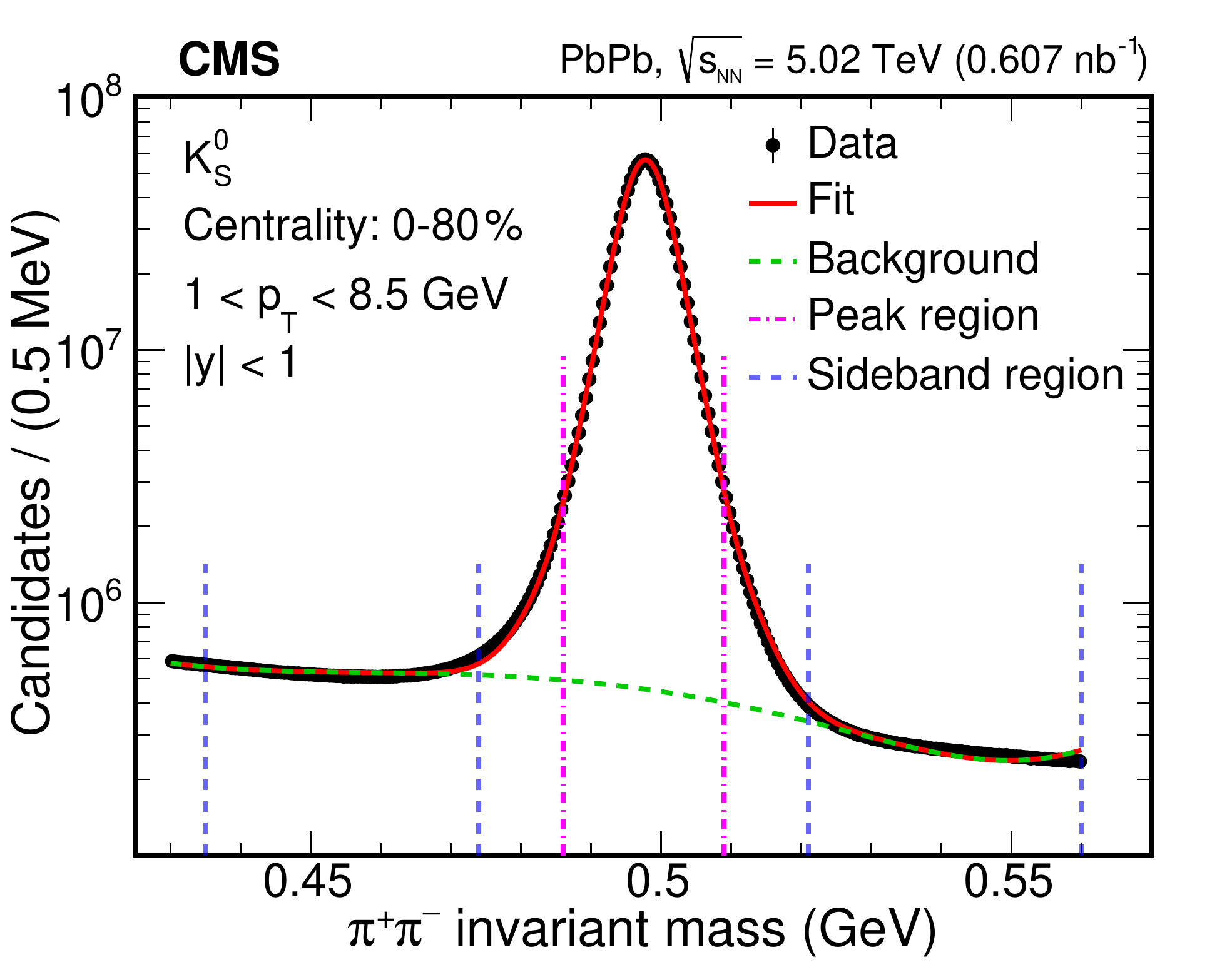}
 \includegraphics[width=0.49\textwidth]{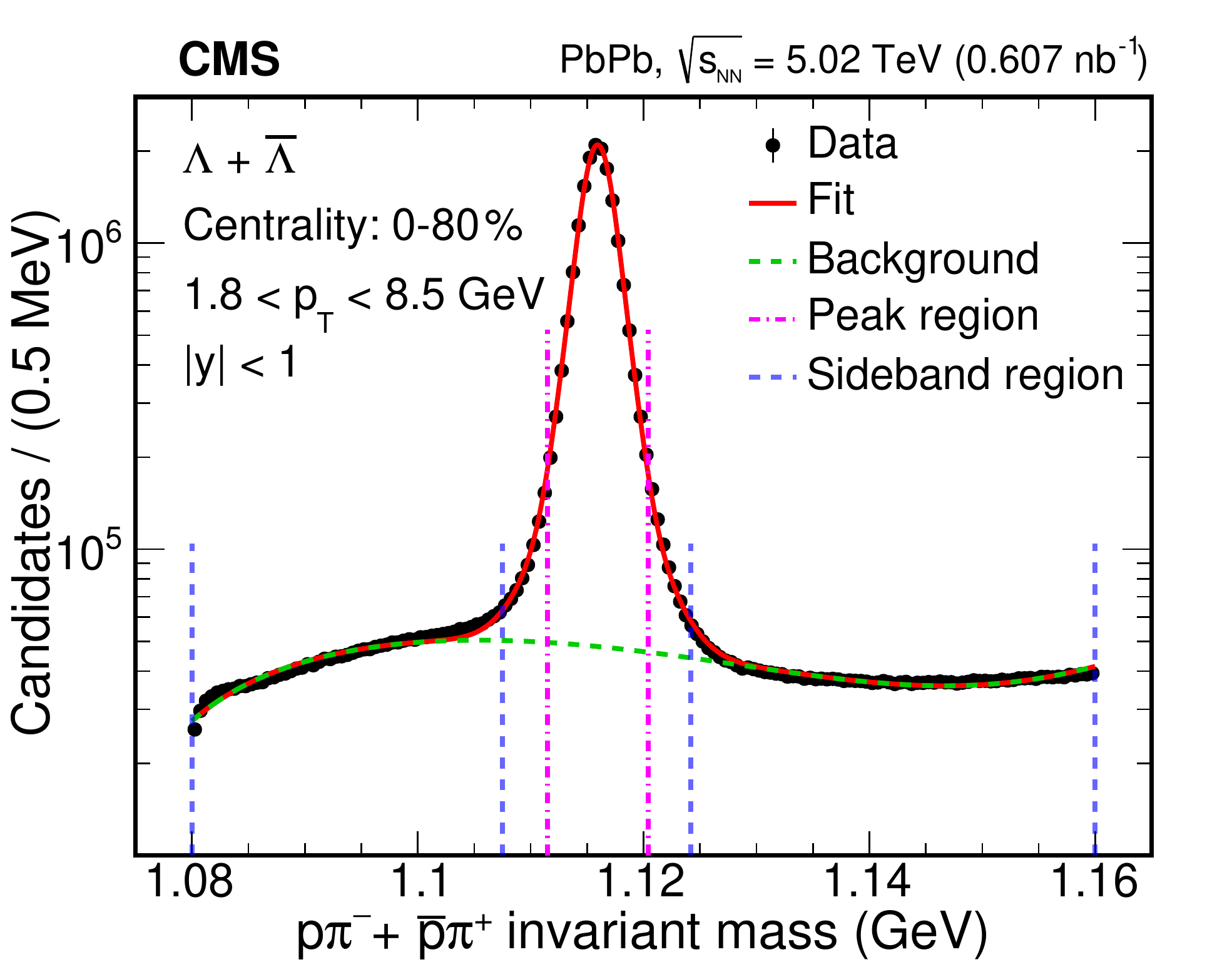}
  \caption{The invariant mass of \KS (left) and \LM (right), and their corresponding fits in the 0--80\% centrality range. The circles are the data, and the fit is shown with a solid (red) line for the total fit, and a dashed (green) line for the background fit. The vertical dashed-dotted (pink) lines indicate the peak region and the vertical dashed (blue) lines indicate the sideband regions.}
 \label{fig:KS_LM_ALM_mass}
 \end{figure*}

Fits to the invariant mass spectrum are performed using a sum of three Gaussian functions with a common mean to describe the signal distribution and a fourth-order polynomial to describe the background. These empirical functions were chosen to provide a good description of the data. Peak and sideband invariant mass regions are defined to select events dominated by signal and background, respectively. Defining  $\sigma$ as the average resolution based on the Gaussian sum, the peak regions are selected to be within ${\pm}2\sigma$ from the nominal \VZ mass and are given by $486 < \mathrm{M(\PGpp\PGpm)} < 509\MeV$ and $1111.5 < \mathrm{M(\Pp\PGpm)} < 1120.4\MeV$ for \KS and \LM candidates, respectively. The sideband regions are selected to be more than $4\sigma$ from the nominal \VZ mass and are given by $23.5 < \abs{\mathrm{M(\PGpp\PGpm)} - 497.5} < 62.5\MeV$ for \KS candidates and $1080 < \mathrm{M(\Pp\PGpm)} < 1107.5\MeV$ together with $1124.2 < \mathrm{M(\Pp\PGpm)} < 1160\MeV$ for \LM candidates. Examples of invariant mass distributions for $\PGpp\PGpm$ and $\Pp\PGpm$ pairs, and their corresponding fits in the 0--80\% centrality range, are shown in Fig.~\ref{fig:KS_LM_ALM_mass}.

 \section{Analysis method}
 \label{sec:ana}
 The two-particle correlation is constructed as
 
 \begin{equation}
 \cobs = \mathcal{N}\frac{\aobs}{\bobs},
 \label{cooreqwop}
 \end{equation}
 where \cobs is the observed normalized pair yield as a function of the invariant relative momentum \qinv in the pair rest frame, defined as~\cite{Lisa:2005dd}
  \begin{equation}
         \begin{split}
                 & \qinv =  \sqrt{{-Q^{\mu}Q_{\mu}}}, \\
                 & Q^{\mu} = (k_1 - k_2)^{\mu} - \frac{(k_1 - k_2)^{\mu} P_{\mu}}{P^{\mu}P_{\mu}}P^{\mu},
         \end{split}
 \end{equation}
 
 \noindent where  $P = k_1 + k_2$, and $k_1$ and $k_2$ are the four momenta of the \VZ particles. Note that for two particles of the same mass, the second term of $Q^{\mu}$ is zero.

The distribution \aobs is the signal distribution, which includes the femtoscopic and nonfemtoscopic correlations formed by pairing the selected \VZ particles from a given event. The reference distribution \bobs is constructed by mixing the \VZ particles from different events, as described in Ref.~\cite{KOPYLOV1974472}. In this procedure, the \VZ particle from one event is paired with \VZ particles from 30 different events. To ensure that the 30 events used in the mixing are similar to the signal events, the centrality and primary vertex of each mixed event must be within 5\% and 2\unit{cm}, respectively, of those in the corresponding signal event. By dividing \aobs by \bobs, CMS detector and reconstruction artifacts are strongly suppressed. The normalization factor $\mathcal{N}$ is the ratio of the number of pairs in the reference distribution to that in the signal distribution. Because of the background in the peak region of the invariant mass distributions, the measured signal distribution $(\aobs)$ contains contributions from signal-signal $(\ass)$, signal-background $(\asb)$, and background-background $(\abb)$ correlations. The measured \aobs distribution can be written as

\ifthenelse{\boolean{cms@external}}
{ 
\begin{multline}
\aobs = \fss\ass + \fsb\asb\\
+ \fbb\abb.
\label{purityeq_1st}
\end{multline}
}
{
\begin{equation}
\aobs = \fss\ass + \fsb\asb+ \fbb\abb.
\label{purityeq_1st}
\end{equation}
}
The distributions \asb and \abb are obtained from the peak-sideband and sideband-sideband combinations, respectively. The small amount of background (signal) contamination in the signal (sideband) region has a negligible effect on the shape of \asb $(\abb)$. All distributions, \aobs, \asb, and \abb are normalized to unity. The parameters, \fss, \fsb, and \fbb are the signal-signal, signal-background, and background-background fractions, extracted using an invariant mass fit based on combinatorial analyses with
 \begin{equation}
         \begin{split}
                 & \fss = \frac{\binom{\mathrm{s}}{2}}{\binom{\mathrm{s}+\mathrm{b}}{2}},\\
                 & \fbb = \frac{\binom{\mathrm{b}}{2}}{\binom{\mathrm{s}+\mathrm{b}}{2}}, \;\text{and} \\ 
                 & \fsb = 1-f^{\mathrm{ss}}-f^{\mathrm{bb}},
         \end{split}
 \end{equation}
 where $\binom{n}{2}=\frac{n(n-1)}{2}$ is the binomial coefficient, which returns the number of ways that a pair can be chosen from $n$ objects. The quantities $\mathrm{s}$ and $\mathrm{b}$ in the binomial coefficients are the number of signal and background particles, respectively, obtained by integrating the appropriate function from the fit to the invariant mass distribution. The resulting values of the \fss, \fsb, and \fbb parameters are presented in Table~\ref{tab:ppfraction}.
\begin{table*}[ht]\renewcommand{\arraystretch}{1.2}
\topcaption{The values of the \fss, \fsb, and \fbb parameters are based on the number of counts in the peak and sideband regions. The values are given for the \KSKS, \LALKS, and \LLALAL pairs in the integrated 0--80\% centrality range. The uncertainties are statistical.}
\centering
\begin{tabular}{ c  c  c  c }
Parameter  & \KSKS & \LALKS & \LLALAL \\
\hline
{\fss} & $0.9569 \pm 0.0006$ & $0.9332 \pm 0.0006$ & $0.9054 \pm 0.0020$ \vspace{0.12cm}\\
{\fsb} & $0.0426 \pm 0.0005$ & $0.0655 \pm 0.0006$ & $0.0922 \pm 0.0018$ \vspace{0.12cm}\\ 
{\fbb} & $0.0005 \pm 0.0001$ & $0.0013 \pm 0.0002$ & $0.0024 \pm 0.0002$ \vspace{0.12cm}\\ 
\end{tabular}
\label{tab:ppfraction}
\end{table*}

Once we have all the distributions $(\aobs$, $\asb$, and $\abb)$ and the parameters $(\fss$, $\fsb$, and $\fbb)$, the \ass distribution can be extracted using Eq.~(\ref{purityeq_1st}), with
\ifthenelse{\boolean{cms@external}}
{
\begin{multline}
\ass = \Big[\aobs - \fsb(\asb)\\
- \fbb(\abb)\Big]/\fss.
\end{multline}
}
{
\begin{equation}
\ass = \Big[\aobs - \fsb(\asb)- \fbb(\abb)\Big]/\fss.
\end{equation}
}
\label{purityeq}
The same procedure is followed for the reference distribution. After extracting the \ass and \bss distributions, the correlation distribution is calculated as
\begin{equation}
 \css = \mathcal{N}\frac{\ass}{\bss}.
 \label{ppcorr}
 \end{equation}
The measured fractions, reported in Table~\ref{tab:ppfraction}, confirm the high purity of the \VZ pairs suggested by Fig.~\ref{fig:KS_LM_ALM_mass}. The effect of the non-\VZ background correction on the measured physics parameters is small compared to either the statistical or systematic uncertainties. While the \css distribution is corrected for non-\VZ backgrounds as well as nonuniform efficiency and acceptance, it still includes nonfemtoscopic background correlations, such as those associated with elliptic flow~\cite{Kisiel:2017gip}, minijets~\cite{PhysRevC.97.064912}, resonance decays~\cite{PhysRevC.97.064912}, and energy-momentum conservation~\cite{PhysRevC.99.024001}. The nonfemtoscopic background contribution is modeled using an empirically determined double Gaussian function
\begin{equation}
\Omega(\qinv)=N\left(1 + \alpha_1\re^{-\qinv^{2}R^2_1}\right) \left(1 - \alpha_2\re^{-\qinv^{2}R^2_2}\right),
\label{clustereq}
\end{equation}
where $N$, $\alpha_1$, $\alpha_2$, $R_1$, and $R_2$ are fit parameters. This function was selected for its reproduction of the distributions in both real data at high \qinv and simulated data that do not include the correlations being measured.

Fits are performed to the \css distributions to extract the source size and strong interaction scattering parameters. As the \VZ particles are neutral, the Coulomb interaction is absent. However, the correlations are sensitive to QS and FSI effects, with $s$-wave interactions assumed to dominate for the small relative momenta of the particle pairs analyzed. The correlation distribution for all pairs (\KSKS, \LALKS, and \LLALAL) is interpreted in the LL model. This model relates the two-particle correlation function to the source size and also takes into account FSI effects~\cite{Lednicky:1981su}. The general correlation function is
\ifthenelse{\boolean{cms@external}}
{ 
\begin{multline}
C_{\text{total}}(\qinv) = \Big[1+ \Lam \Big(\cqs\\
 + \cfsi\Big)\Big]\Omega(\qinv),
\label{totalcorr}
\end{multline}
}
{
\begin{equation}
C_{\text{total}}(\qinv) = \Big[1+ \Lam \Big(\cqs + \cfsi\Big)\Big]\Omega(\qinv),
\label{totalcorr}
\end{equation}
}
where \cqs is the QS function and \cfsi is the FSI function. The parameter \Lam is referred to as the correlation strength. In the absence of FSI effects, \Lam equals unity for a perfectly incoherent Gaussian source. However, effects such as resonance decay, minijets, coherent sources, non-Gaussian source contributions, and FSI between particles can lead to deviations of the \Lam parameter from unity.

Neglecting CP violation, the \KSKS system can be written as 
\ifthenelse{\boolean{cms@external}}
{ 
\begin{multline}
\ket{\KSKS} = \frac{1}{2}\Bigl(\ket{\PKZ\PKZ} + \ket{\PKZ\PKAZ} + \ket{\PKAZ\PKZ}\\
 + \ket{\PKAZ\PKAZ}\Bigr). 
\label{ksksstate}
\end{multline}
}
{
\begin{equation}
\ket{\KSKS} = \frac{1}{2}\Bigl(\ket{\PKZ\PKZ} + \ket{\PKZ\PKAZ} + \ket{\PKAZ\PKZ} + \ket{\PKAZ\PKAZ}\Bigr).
\label{ksksstate}
\end{equation}
}
It can be shown~\cite{Lednicky:1981su, PhysRevC.74.054902} that the resulting correlations follow Bose--Einstein quantum statistics, with
\begin{equation}
\cqs = \re^{(-\qinv^{2}\Rin^{2})},
\label{KSqs}
\end{equation}
where the source radius \Rin reflects the size of the region over which the particles are emitted.

The FSI for the \KSKS correlations is modeled by~\cite{Lednicky:1981su, PhysRevC.74.054902}
\ifthenelse{\boolean{cms@external}}
{ 
\begin{multline}
\cfsi = \frac{1}{2} \Biggl[\left|\frac{f(k)}{\Rin}\right|^2+\frac{4\Re f(k)}{\sqrt{\pi}\Rin}F_1(\qinv\Rin)\\
-\frac{2\Im f(k)}{\Rin}F_2(\qinv\Rin)\Biggr],
\end{multline}
}
{
\begin{equation}
\cfsi = \frac{1}{2} \Biggl[\left|\frac{f(k)}{\Rin}\right|^2+\frac{4\Re f(k)}{\sqrt{\pi}\Rin}F_1(\qinv\Rin)-\frac{2\Im f(k)}{\Rin}F_2(\qinv\Rin)\Biggr],
\end{equation}
}
\label{KSfsi}
where
 \begin{equation}
    \begin{split}
      & k = \qinv/2, \\ 
      & F_1(z)=\frac{1}{z}\re^{-z^2}\int_0^z e^{x^2} \rd{x}, \;\text{and} \\
      & F_2(z) = \frac{1-\re^{-z^2}}{z}. 
      \end{split}
    \end{equation}
The function $f(k)$ is the $\PKZ\PKAZ$ $s$-wave scattering amplitude, with real and imaginary parts $\Re f(k)$ and $\Im f(k)$, respectively. This amplitude is dominated by the near-threshold $s$-wave isoscalar resonance $\PfDzP{980}$ and the $s$-wave isovector resonance \Paz, with the total scattering amplitude given by an average of these contributions: $f(k)=(f_{\PfDzP{980}}(k)+f_{\Paz}(k))/2$. The individual resonance amplitudes depend on the resonance mass $m_r$, with $r = \PfDzP{980}$ or \Paz, the kaon mass $m_\PK$, and the resonance couplings $\gamma_r$ ($\gamma'_r$) to the $\PKZ\PKAZ$ $(\PGp\PGp$ for $\PfDzP{980}$ and $\PGpz\PGh$ for $\Paz)$ channels. Then, $f_r(k) = \gamma_r/\!\left[m^2_r-\zeta-i\gamma_rk-i\gamma'_rk'_r\right]$, where $\zeta=4(m^2_\PK+k^2)$ and $k'_r$ denotes the momentum in the second ($\PGp\PGp$ or $\PGpz\PGh$) decay channel with the corresponding partial width $\Gamma' = \gamma'_rk'_r/m_r$ (more details can be found in Ref.~\cite{PhysRevC.74.054902}). The scattering amplitude is calculated using the resonance mass and the coupling parameters from Refs.~\cite{Martin:1979520,Antonelli:2002ip,Achasov:2001cj,Achasov:2003}, taken from row C of Table~1 of Ref.~\cite{PhysRevC.74.054902}.

For the correlations involving \LM baryons, the \cqs and \cfsi functions are~\cite{Lednicky:1981su}
\ifthenelse{\boolean{cms@external}}
{ 
 \begin{multline}
   \quad\,\cqs = \alpha \re^{(-\qinv^{2}\Rin^{2})},\quad\text{and}\\
    \cfsi = (1+\alpha)\Bigg[\frac{1}{2}\frac{|f(k)|^2}{\Rin^{2}}\big(1-\frac{1}{2\sqrt{\pi}}\frac{d_0}{\Rin}\big)\\
      +\frac{2\Re{f(k)}}{\sqrt{\pi}\Rin}F_1(\qinv\Rin)-\frac{\Im{f(k)}}{\Rin}F_2(\qinv\Rin)\Bigg],
    \label{LMqsfsi}
 \end{multline}
}
{
\begin{equation}
\begin{split}
  &\cqs = \alpha \re^{(-\qinv^{2}\Rin^{2})},\quad\text{and}\\
  &\cfsi = (1+\alpha)\Bigg[\frac{1}{2}\frac{|f(k)|^2}{\Rin^{2}}\big(1-\frac{1}{2\sqrt{\pi}}\frac{d_0}{\Rin}\big)+\frac{2\Re{f(k)}}{\sqrt{\pi}\Rin}F_1(\qinv\Rin)-\frac{\Im{f(k)}}{\Rin}F_2(\qinv\Rin)\Bigg],
\end{split}
\label{LMqsfsi}
\end{equation}
}
where $\alpha = {-}1/2$ for \LLALAL correlations for two identical fermions and  $\alpha = 0$ for \LALKS correlations as there are no QS effects for nonidentical particles~\cite{Lednicky:1981su}. The scattering amplitude $f(k)$ is parameterized by a complex scattering length ($f_0$) and an effective range ($d_0$) with $f(k) = [{1}/{f_0}+d_0k^2/2-ik]^{-1}$~\cite{Lednicky:1981su}. The imaginary part of $f_0$ is responsible for inelastic processes (annihilation). For an attractive interaction that is not strong enough to produce a bound state, the real part of $f_0$ is positive, while a repulsive interaction corresponds to a negative $\Re f_0$ of the order of the range of the repulsive potential. In the presence of a bound state, $\Re f_0$ is also negative, but with a much larger magnitude. The femtoscopic sign convention and notation for the scattering length differ from those used in nuclear physics, where the corresponding scattering length $a_0 = - f_0$. As the \LALKS and \LLALAL correlations each have only one spin state that contributes to the $s$-wave scattering, Eq.~(\ref{LMqsfsi}) suffices to describe the FSI effects.

Fits to the correlation distribution of all the pairs were performed using Eq.~(\ref{totalcorr}) with the nonfemtoscopic background parameters ($N$, $\alpha_1$, $\alpha_2$, $R_1$, and $R_2$) treated as free parameters. For \KSKS correlations, the parameters of interest are \Rin and \Lam, with the scattering amplitude based on previous measurements~\cite{Martin:1979520,Antonelli:2002ip,Achasov:2001cj,Achasov:2003}. The \LALKS and \LLALAL correlations include additional parameters: $d_0$, $\Re f_0$, and $\Im f_0$. The $\Im f_0$ term for \LLALAL correlations is set to zero since there are no baryon-baryon annihilation processes.

\begin{figure*}[htbp]
\centering
\includegraphics[width=0.98\textwidth]{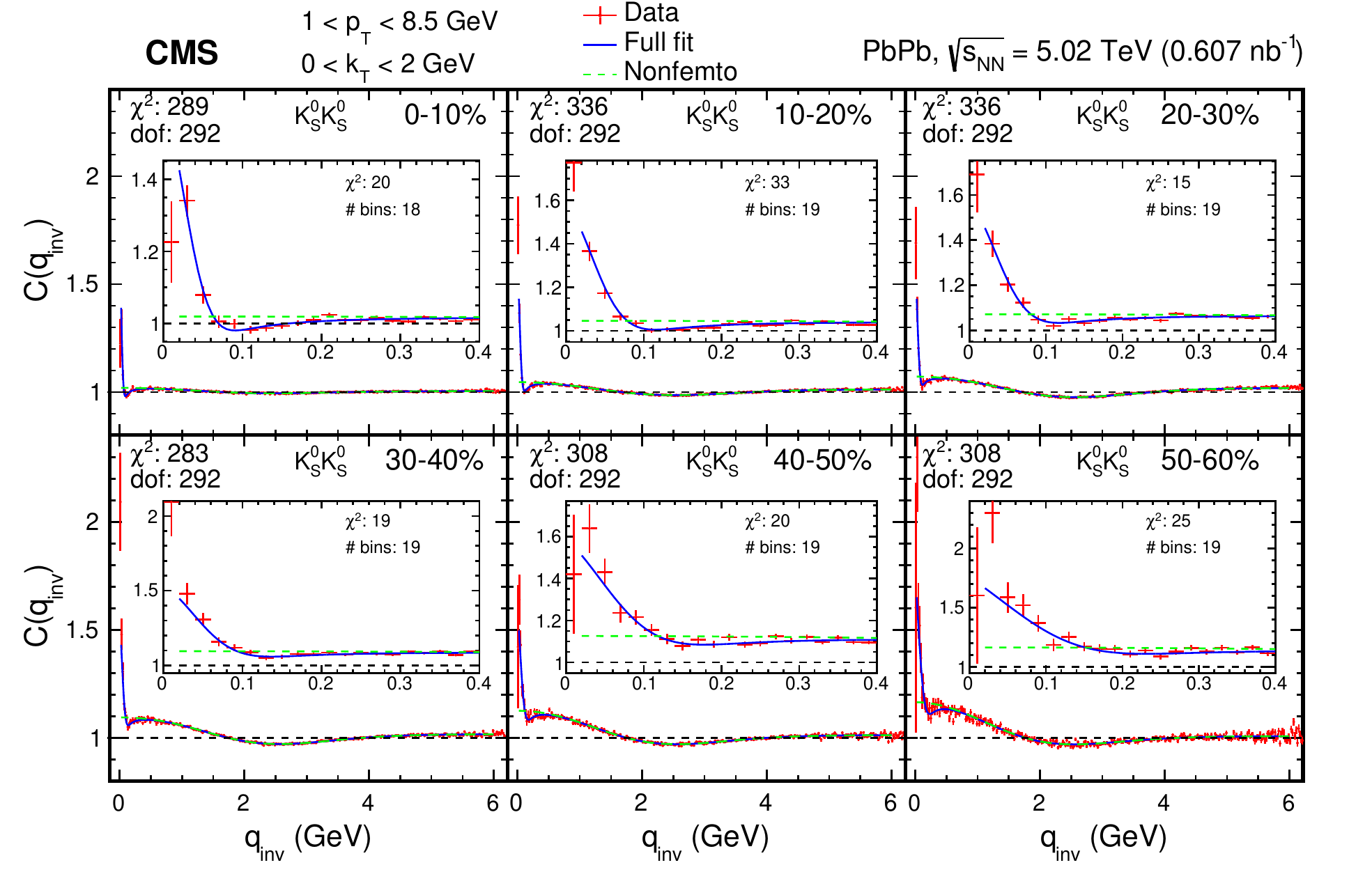}
	\caption{The correlation distributions and fits for \KSKS pairs in different centrality ranges, starting from 0--10\% centrality to 50--60\% centrality, with $0 < \kt < 2 \GeV$. In each plot, the red circles are the data, the blue solid line is the fit using Eq.~(\ref{totalcorr}), and the green dashed line is the nonfemtoscopic background from Eq.~(\ref{clustereq}). The $\chi^2$ and $\mathrm{dof}$ values are for the full \qinv range. The insert plots show the data and the fit for the $\qinv < 0.4\GeV$ region, with the $\chi^2$ and number of bins evaluated in that region.}
\label{fig:ks_corr_meas}
\end{figure*}

\begin{figure*}[htbp]
\centering
\includegraphics[width=0.49\textwidth]{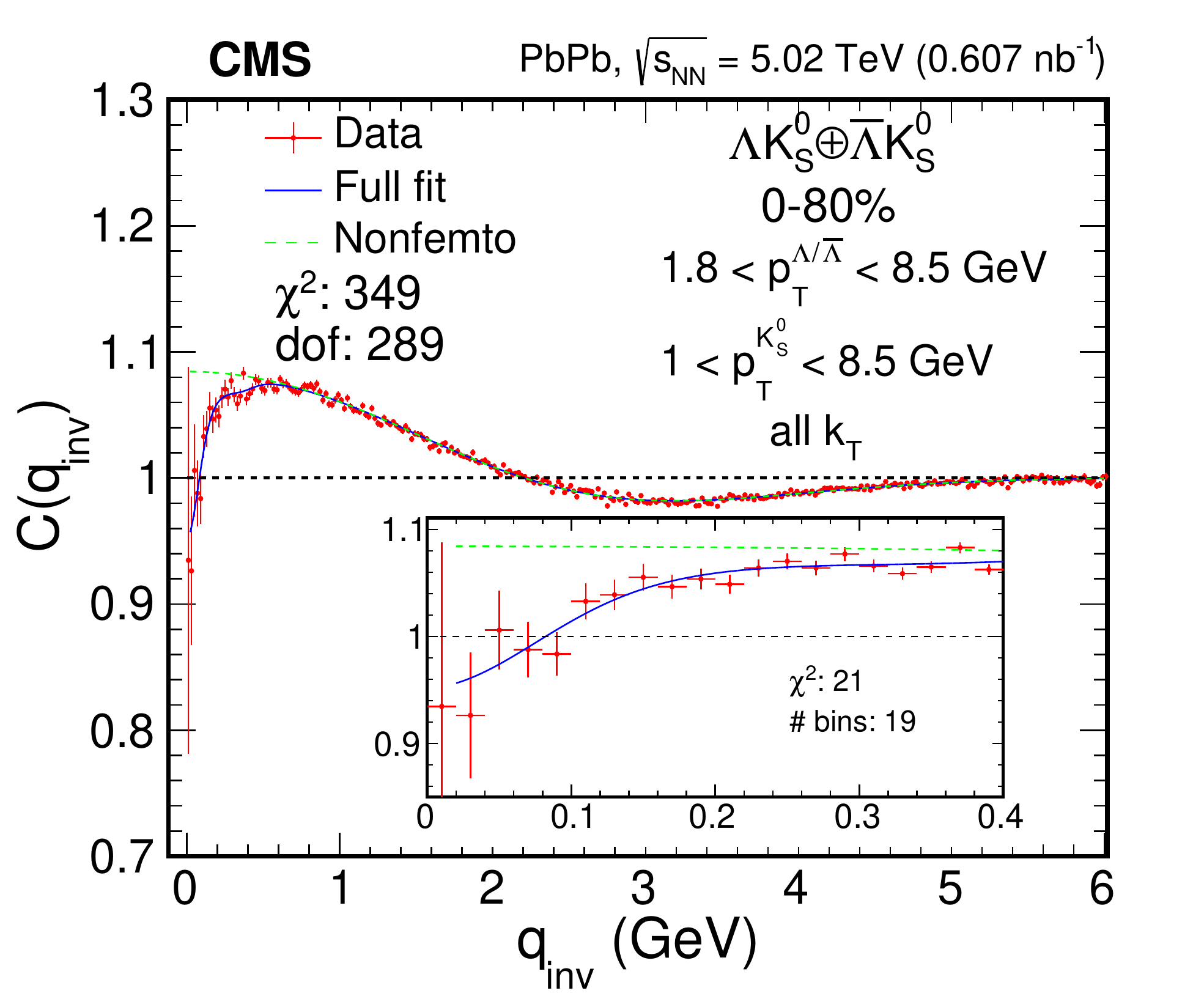}
\includegraphics[width=0.49\textwidth]{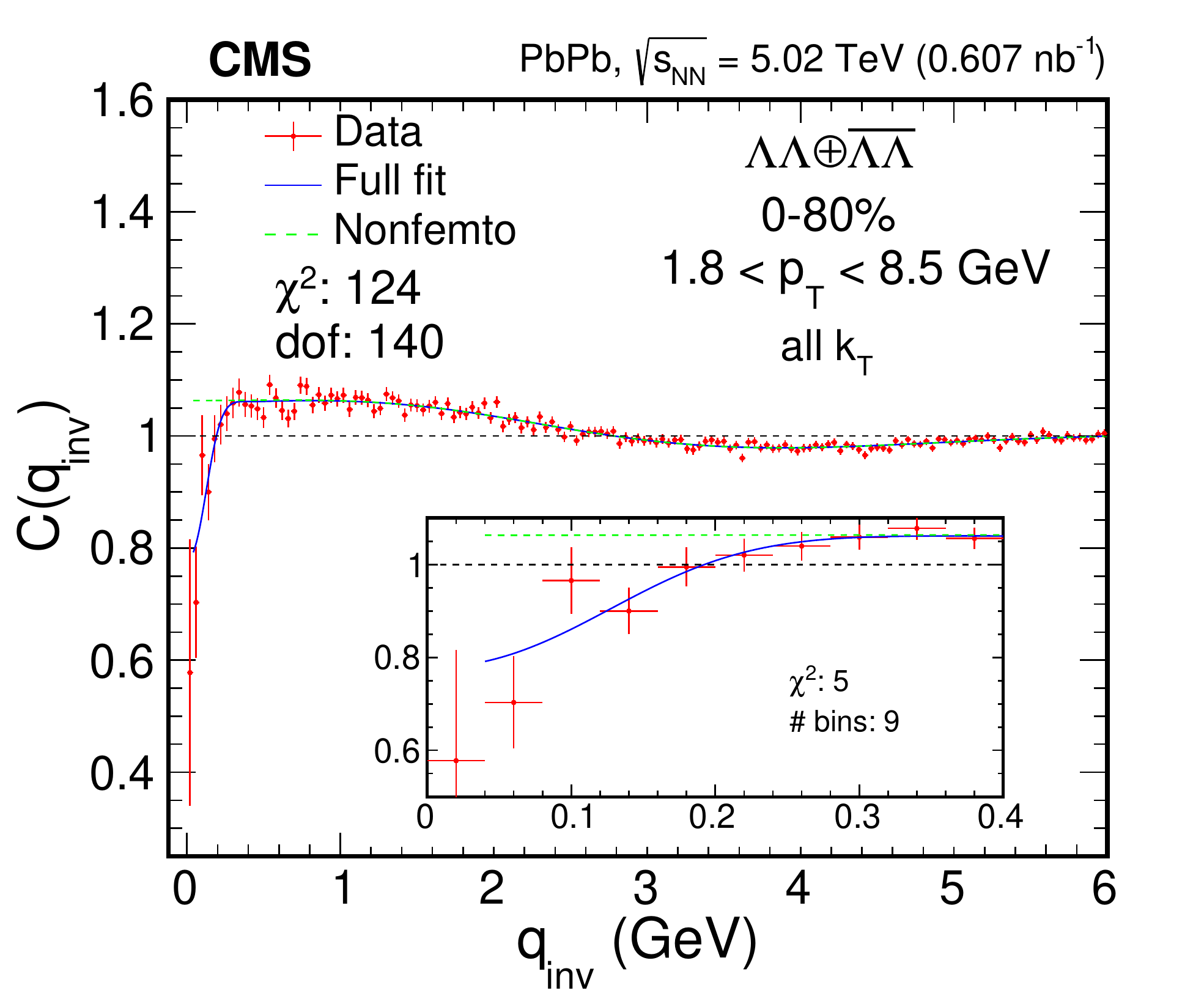} 
	\caption{The correlation distributions and fits for \LALKS (left) and \LLALAL (right) pairs with 0--80\% centrality and no restriction on \kt. In each plot, the red circles are the data, the blue solid line is the fit using Eq.~(\ref{totalcorr}), and the green dashed line is the nonfemtoscopic background from Eq.~(\ref{clustereq}). The $\chi^2$ and $\mathrm{dof}$ values are for the full \qinv range. The insert plots show the data and the fit for the $\qinv < 0.4\GeV$ region, with the $\chi^2$ and number of bins evaluated in that region.}
\label{fig:lks_ll_corr_meas}
\end{figure*}

Histograms of the correlation distributions are generated in the range $0 < \qinv < 6\GeV$ with 0.02\GeV wide bins for the \KSKS and \LALKS correlations and 0.04\GeV wide bins for the \LLALAL correlations. The fits exclude the first \qinv bin to avoid a potential bias from the method used to address cases where \VZ candidates share daughter tracks. Studies using simulated events indicate that only this first bin is affected by this remediation. Least-square fits are performed to the experimental data with the uncertainties in the fit parameters calculated using the \textsc{minos} technique~\cite{James:1975dr}. Examples of correlation measurements and their fits and corresponding $\chi^2/\mathrm{dof}$ values, are presented in Figs.~\ref{fig:ks_corr_meas} and~\ref{fig:lks_ll_corr_meas}. The \KSKS correlations, shown in Fig.~\ref{fig:ks_corr_meas}, are independently fitted for each of the six centrality bins with $0 < \kt < 2\GeV$, where $\kt\equiv\abs{\vec{p}_{\mathrm{T},1}+\vec{p}_{\mathrm{T},2}}/2$ is the average transverse momentum of the particle pair. The \LALKS (left) and \LLALAL (right) correlations, shown in Fig.~\ref{fig:lks_ll_corr_meas}, involve fewer events and, therefore, only a single fit is performed for each, with the data integrated over the centrality range 0--80\% and with no restriction on \kt.

Using events simulated with the \HYDJET and \HIJING 1.3~\cite{HIJING2} event generators, we find that 85\% and 39\% of the selected \LM baryons are directly produced, respectively. Electromagnetic \PgSz baryon decays account for 8\% and 13\%, respectively, while weak decays of \PgX/\PgO baryons account for 7\% and 48\% of the observed yields. Given the wide variation in the feed-down contribution and the lack of experimental measurements for these contributions, our results involving \LM baryons are presented without being corrected for potential, and possibly significant, feed-down effects and residual correlations.

\section{Systematic uncertainties}
\label{subsec:app_sysun}
The systematic uncertainties for the fit parameters are based on the changes found in the parameter values after individually varying each of the analysis criteria, as discussed below. In cases with more than one variation for a single source, the maximum deviation from the nominal value is used. The total systematic uncertainty is obtained by adding the uncertainties from each source in quadrature. The BDT discriminant is varied so as to adjust the signal-to-background ratio, with the signal yield changing by $\pm$15\% in the process. The nominal method to account for \VZ candidates sharing daughter tracks is to remove one of the \VZ candidates at random, which is then not used by any pair. Two alternative approaches are used, one in which both \VZ candidates are removed and another in which, for events with multiple \VZ candidate pair combinations, only the pairs in which the two particles share a daughter are removed. The systematic uncertainties related to \VZ signal and background modeling are investigated by varying the background shape from a fourth- to a third-order polynomial and the signal shape from a sum of three Gaussian functions to a sum of two or four Gaussian functions. An alternative nonfemtoscopic background function $\Omega(\qinv)= N(1 + B\re^{-|\qinv/\sigma|^2}) (1 + \epsilon\qinv)$ is used to assess the uncertainty associated with the choice of the nonfemtoscopic background function. The selection requirements used to construct the mixed event sample are varied to require centrality matching of 3\% instead of the nominal 5\% and the primary vertex position matching with 1\unit{cm} instead of the nominal 2\unit{cm}. The peak region requirement is changed from ${<}2.0\sigma$ to ${<}1.5\sigma$ and ${<}2.5\sigma$ and the sideband region selection from ${>}4.0\sigma$ to ${>}3.5\sigma$ and ${>}4.5\sigma$. The upper limit of the correlation distributions fit range is changed to 5 and 7 GeV, and the lower limit is changed to include the first bin. The possible influence of momentum resolution on the calculated parameters has been studied following the procedure described in Ref.~\cite{ALICE:2015hvw}, with an associated systematic uncertainty quoted in Table~\ref{tab:sysuntab_LALKSKSKS}. For the \KSKS correlations, an additional systematic uncertainty is found by varying the mass and coupling parameters for the $\PfDzP{980}$ and \Paz resonances based on values found in rows A, B, and D of Table 1 of Ref.~\cite{PhysRevC.74.054902}. The systematic uncertainties are summarized in Table~\ref{tab:sysuntab_LALKSKSKS}. 

Dividing the signal distribution by the reference distribution is expected to correct for efficiency-dependent effects in the correlations. A variety of checks were performed with simulated samples to confirm this assumption. It was verified that the reconstruction efficiency of a pair of \VZ particles is well described by the product of the efficiencies for each \VZ, so the efficiency of a pair of \VZ particles from mixed events should be the same as from the same event. It was also verified that the efficiency versus \qinv is the same for both the signal and reference distributions. Therefore, no efficiency correction is applied, and no systematic uncertainty is assessed for the \VZ reconstruction efficiency. For the centrality ranges used in this analysis, the effect of the centrality calibration and resolution is negligible~\cite{PhysRevC.100.044902}. 

\begin{table*}[ht]
\centering
\topcaption{Summary of absolute systematic uncertainties in \KSKS, \LALKS, and \LLALAL correlation measurements. The values for \Rin, $d_0$, $\Re f_0$, and $\Im f_0$ are in\unit{fm}.}
\begin{tabular}{l@{\;}c@{\;\;}c@{}c@{\quad}c@{\;\;}c@{\;\;}c@{\;\;}c@{\;\;}c@{}c@{\quad}c@{\;\;}c@{\;\;}c@{\;\;}c@{}} 

\multirow{2}{*}{Uncertainty source} & \multicolumn{2}{c}{\KSKS} & & \multicolumn{5}{c}{\LALKS} & & \multicolumn{4}{c}{\LLALAL}\\
& {\Rin} & {\Lam} & & {\Rin} & {$d_0$} & {$\Re f_0$} & {$\Im f_0$} & {\Lam} & & {\Rin} & {$d_0$} & {$\Re f_0$} & {\Lam} \\
\hline
{BDT cut} &  0.06--0.19 & 0.01--0.05 & & 0.19 & 0.75 & 0.10 & 0.07 & 0.03 & & 0.06 & 0.43 & 0.05 & 0.31\\   
{Duplicate \VZ removal} & 0.07--0.35 & 0.01--0.05 & & 0.35 & 0.92 & 0.10 & 0.19 & 0.11 & & 0.01 & 1.14 & 0.05 & 0.14\\
{Mass fit function} &  0.00--0.03 & 0.00--0.02 & & 0.09 & 0.05 & 0.01 & 0.03 & 0.03 & & 0.02 & 0.04 & 0.01 & 0.02\\
{Nonfemtoscopic func.} &  0.01--0.13 & 0.01--0.13 & & 0.02 & 0.17 & 0.05 & 0.07 & 0.03 & & 0.02 & 1.02 & 0.14 & 0.93\\
{Reference sample} &  0.02--0.09 & 0.01--0.06 & & 0.22 & 0.48 & 0.12 & 0.12 & 0.03 & & 0.10 & 1.12 & 0.20 & 0.76\\ 
{Peak region} &  0.02--0.17 & 0.00--0.07 & & 0.43 & 0.10 & 0.05 & 0.17 & 0.08 & & 0.22 & 1.21 & 0.08 & 0.35\\      
{Sideband region} &  0.01--0.03 & 0.00--0.01 & & 0.02 & 0.02 & 0.01 & 0.01 & 0.00 & & 0.01 & 0.03 & 0.01 & 0.04\\      
{Fitting range} &  0.01--0.32 & 0.00--0.11 & & 0.20 & 0.18 & 0.03 & 0.08 & 0.04 & & 0.04 & 1.79 & 0.20 & 0.60\\      
{Momentum resolution} &  0.02--0.07 & 0.02--0.08 & & 0.41 & 0.00 & 0.03 & 0.09 & 0.08 & & 0.01 & 0.08 & 0.01 & 0.02\\      
{$\PfDzP{980}$/\Paz param.} &  0.11--0.16 & 0.03--0.05 & & \NA & \NA & \NA & \NA & \NA & & \NA & \NA & \NA & \NA\\ [\cmsTabSkip]
{Total uncertainty} & 0.30--0.51 & 0.10--0.18 & & 0.78 & 1.31 & 0.20 & 0.32 & 0.17 & & 0.25 & 2.91 & 0.33 & 1.43\\
\end{tabular}
\label{tab:sysuntab_LALKSKSKS}
\end{table*}

\section{Results}
\label{sec:results}
\begin{figure*}[htbp]
\centering
\includegraphics[width=0.49\textwidth]{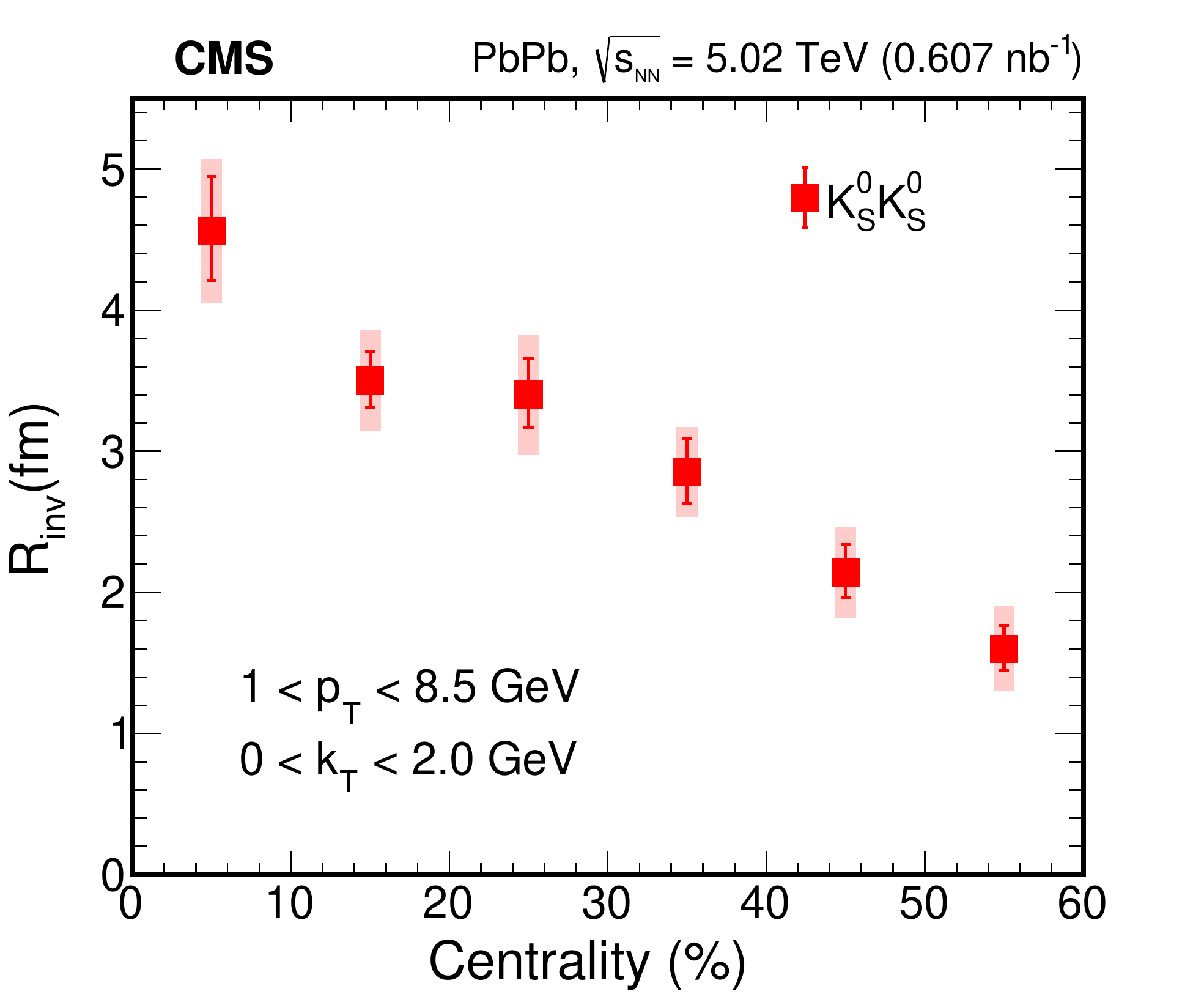}
\includegraphics[width=0.49\textwidth]{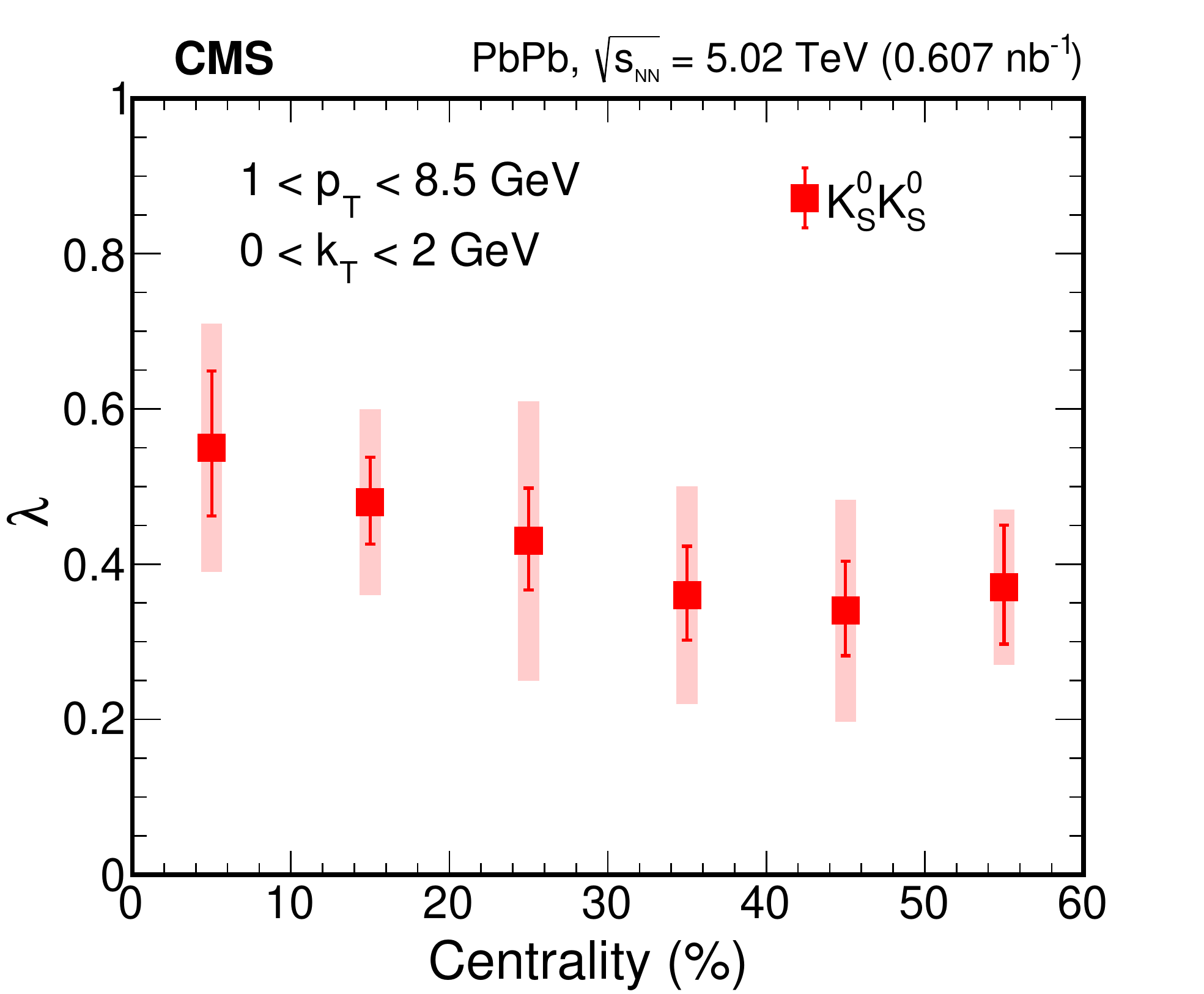}
\caption{The \Rin (left) and \Lam parameter (right) as a function of centrality. For each data point, the lines and the shaded areas indicate the statistical and systematic uncertainties, respectively.} 
\label{fig:rinv_sys}
\end{figure*}

\begin{figure*}[htbp]
\centering
\includegraphics[width=0.49\textwidth]{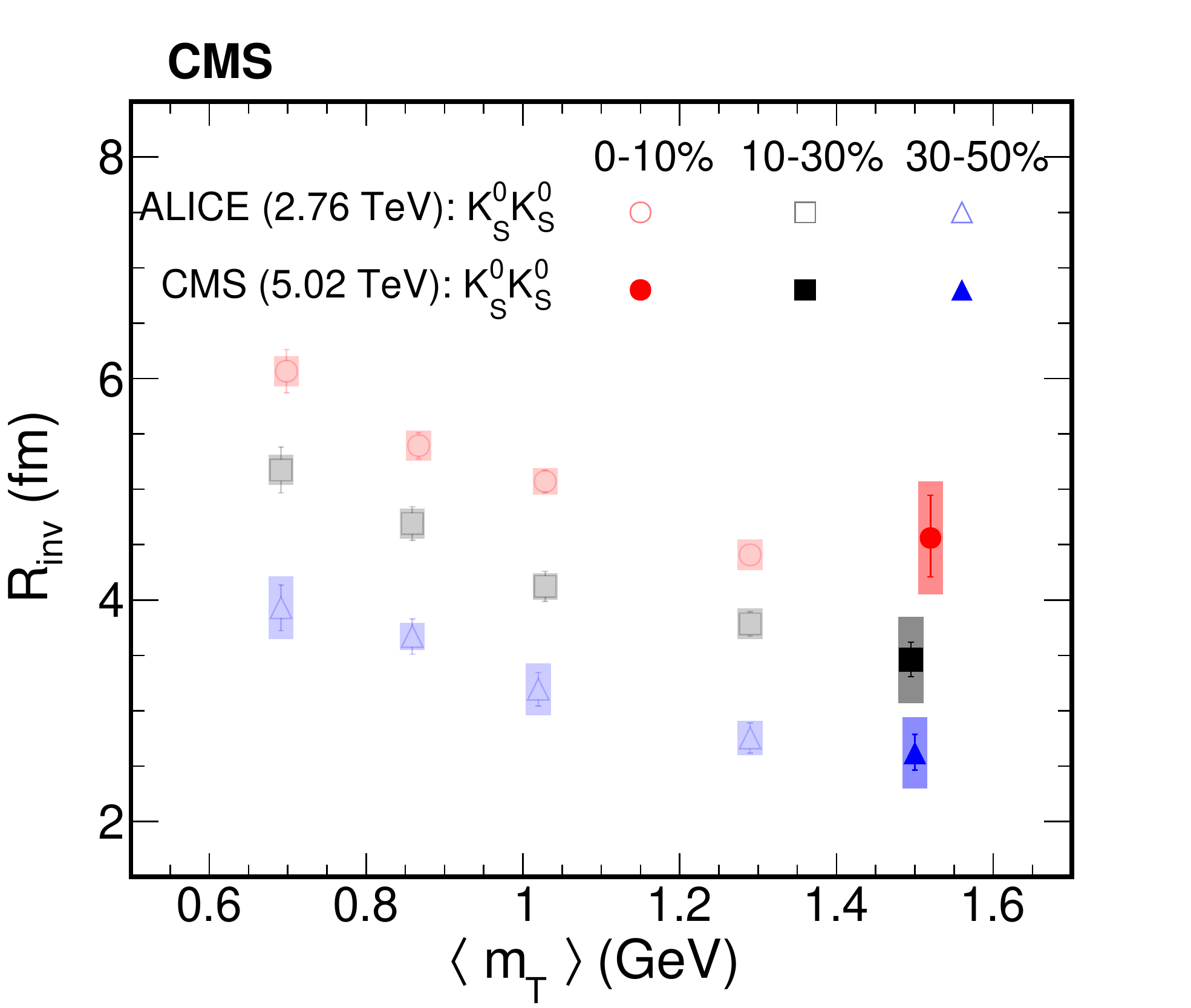}
\includegraphics[width=0.49\textwidth]{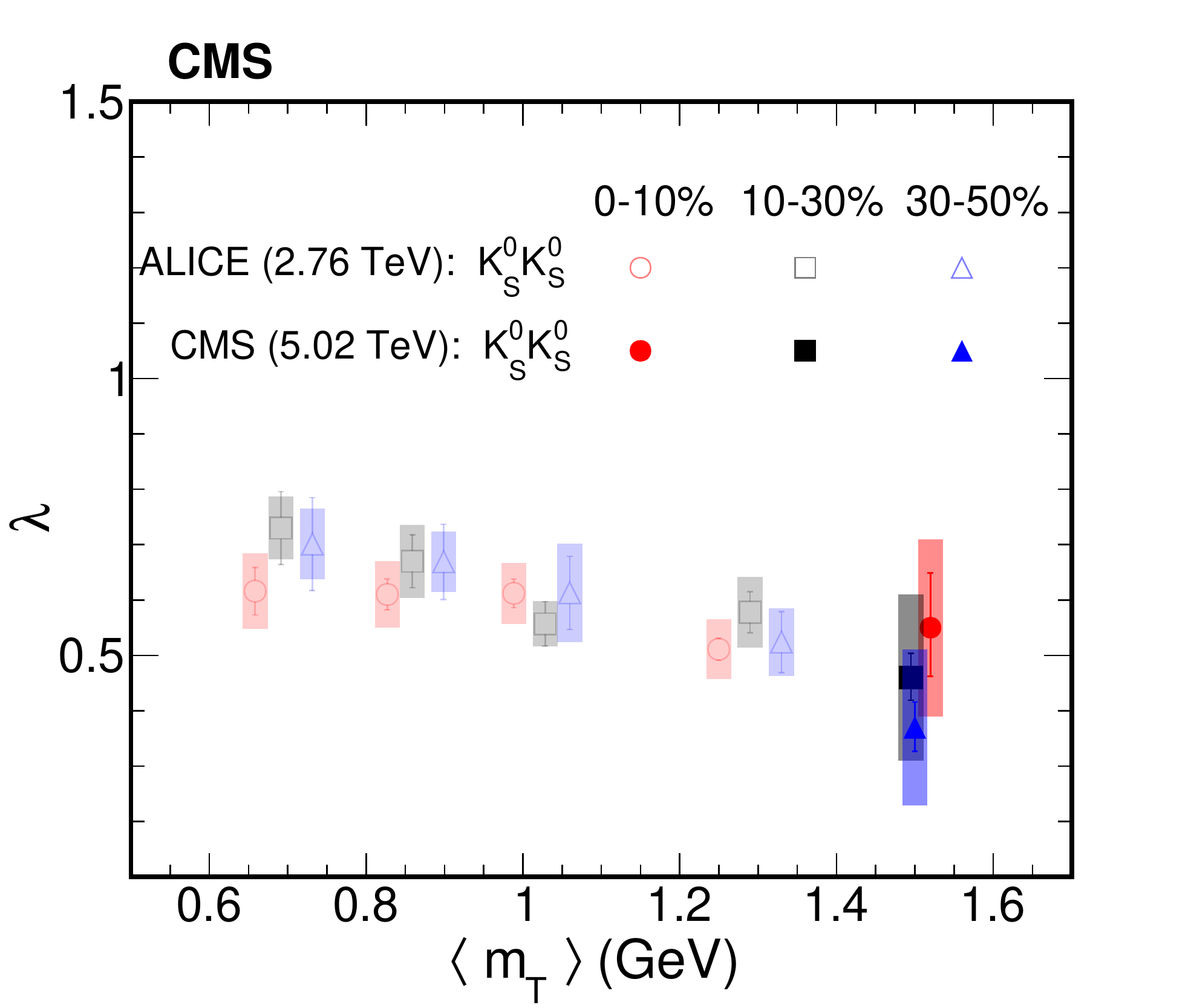}
\caption{The \Rin (left) and \Lam (right) parameters as a function of $\left<\mT\right>$ for the three centrality bins (0--10\%, 10--30\%, and 30--50\%) from this analysis (dark filled symbols) and from the ALICE experiment (light open symbols)~\cite{ALICE:2015hvw}. For each data point, the lines and the shaded areas indicate the statistical and systematic uncertainties, respectively.}
\label{fig:rinv_lambda_CMS_ALICE}
\end{figure*}

The size of the particle emitting source \Rin and the \Lam parameter extracted from the \KSKS correlations for $0 < \kt < 2 \GeV$ are shown as a function of centrality in Fig.~\ref{fig:rinv_sys}. It is observed that the \Rin value decreases from central to peripheral events, as expected from a simple geometric picture of the collisions. Over the full centrality range of 0--80\%, $\Rin = 3.40\pm 0.11\stat\pm 0.37\syst$\unit{fm}. The transverse mass can be calculated as $\mT = \sqrt{\smash[b]{(\minv/2)^2 + \kt^2}}$, where \minv is the invariant mass of the two-particle system~\cite{PhysRevC.103.055201}. The average $\left<\mT\right>$ is evaluated from the transverse mass distribution using two-particle pairs with $\qinv < 0.4\GeV$, accounting for background using the binomial analysis as done for the \qinv distributions, and is found to be 1.50 \GeV and is nearly independent of centrality. Figure~\ref{fig:rinv_lambda_CMS_ALICE} shows CMS and ALICE results for \Rin and \Lam as a function of \mT in three centrality bins. Extrapolations of the \mT dependence found for the ALICE results at $\sqrtsNN = 2.76\TeV$~\cite{ALICE:2015hvw} to the \mT values of the CMS results at $\sqrtsNN = 5.02\TeV$ shows consistent results for the two experiments. The \Lam parameter is seen to decrease from about 0.55 to 0.37 as the collisions become more peripheral. Possible explanations of this decrease include having an increased contribution of minijets and having a source function that becomes increasingly non-Gaussian as the collisions become more peripheral. 

Table~\ref{tab:scatteringpara} includes the extracted \Rin and \Lam parameters as well as $\left<\mT\right>$ for \KSKS, \LALKS, and \LLALAL combinations in the 0--80\% centrality range. A significant decrease is seen in \Rin as the $\left<\mT\right>$ increases. Qualitatively similar results have been found, both for a given pair type in bins of \mT and when comparing multiple pair types~\cite{Lisa:2005dd}. Because of the different minimum \pt requirements for \KS and \LM particles, the variation in $\left<\mT\right>$ includes both $\left<\pt\right>$ and particle mass differences. The anticorrelation of \Rin and $\left<\mT\right>$ has been interpreted as indicating the presence of an expanding source~\cite{Lisa:2005dd}. Although the \Lam parameter obtained from the \LLALAL correlation is consistent, within statistical and systematic uncertainties, with the absence of a femtoscopic correlation, fitting \css under the assumption of $\Lam = 0$ leads to a $\chi^2$ of 32 with 9 bins in the range $0.04 < \qinv < 0.4\GeV$, compared to a $\chi^2$ of 5, with the same number of bins, when the femtoscopic correlation is taken to be present (shown in the right plot of Fig.~\ref{fig:lks_ll_corr_meas}).  

\begin{figure*}[htbp]
\centering
\includegraphics[width=0.49\textwidth]{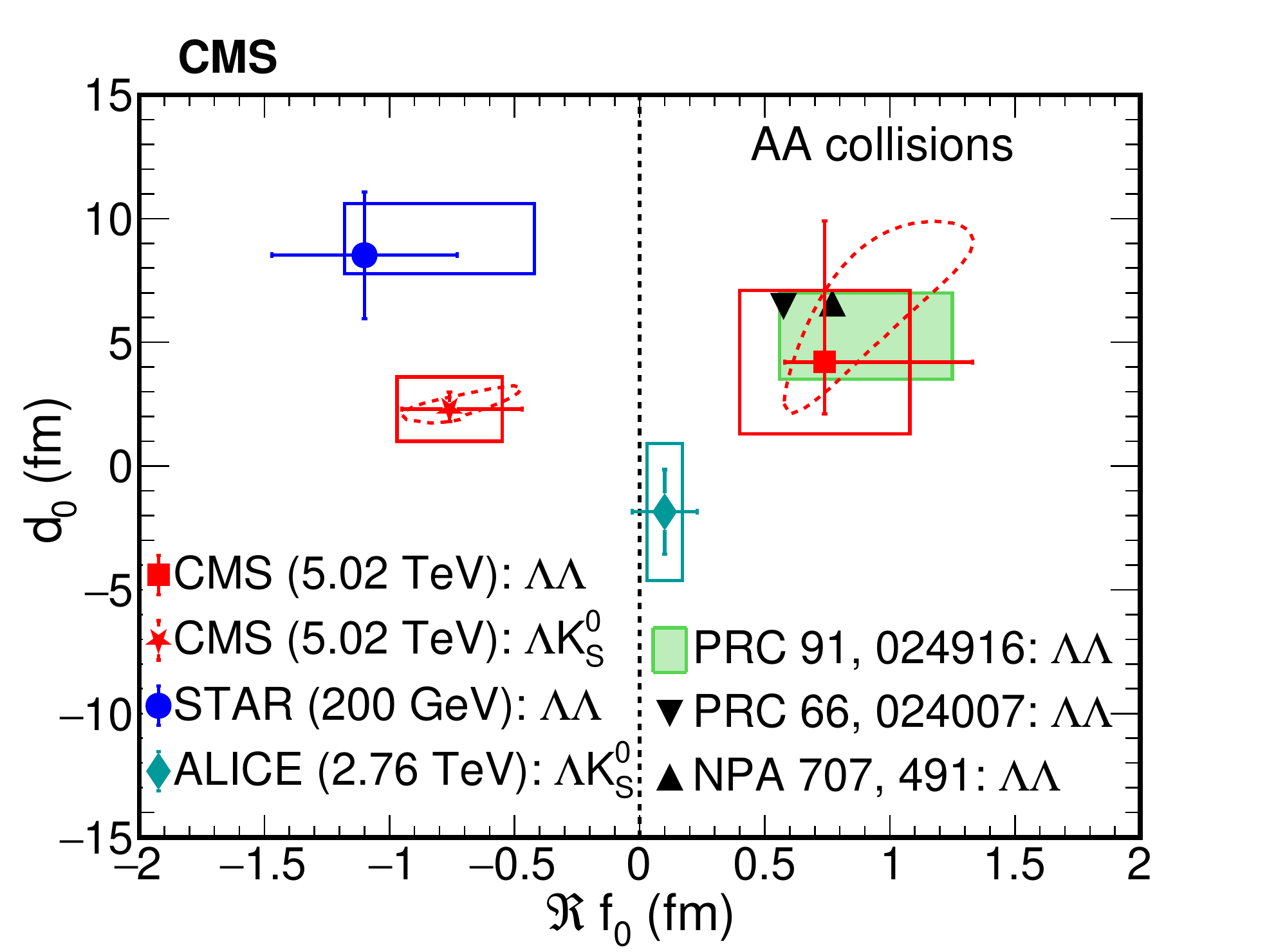}
\includegraphics[width=0.49\textwidth]{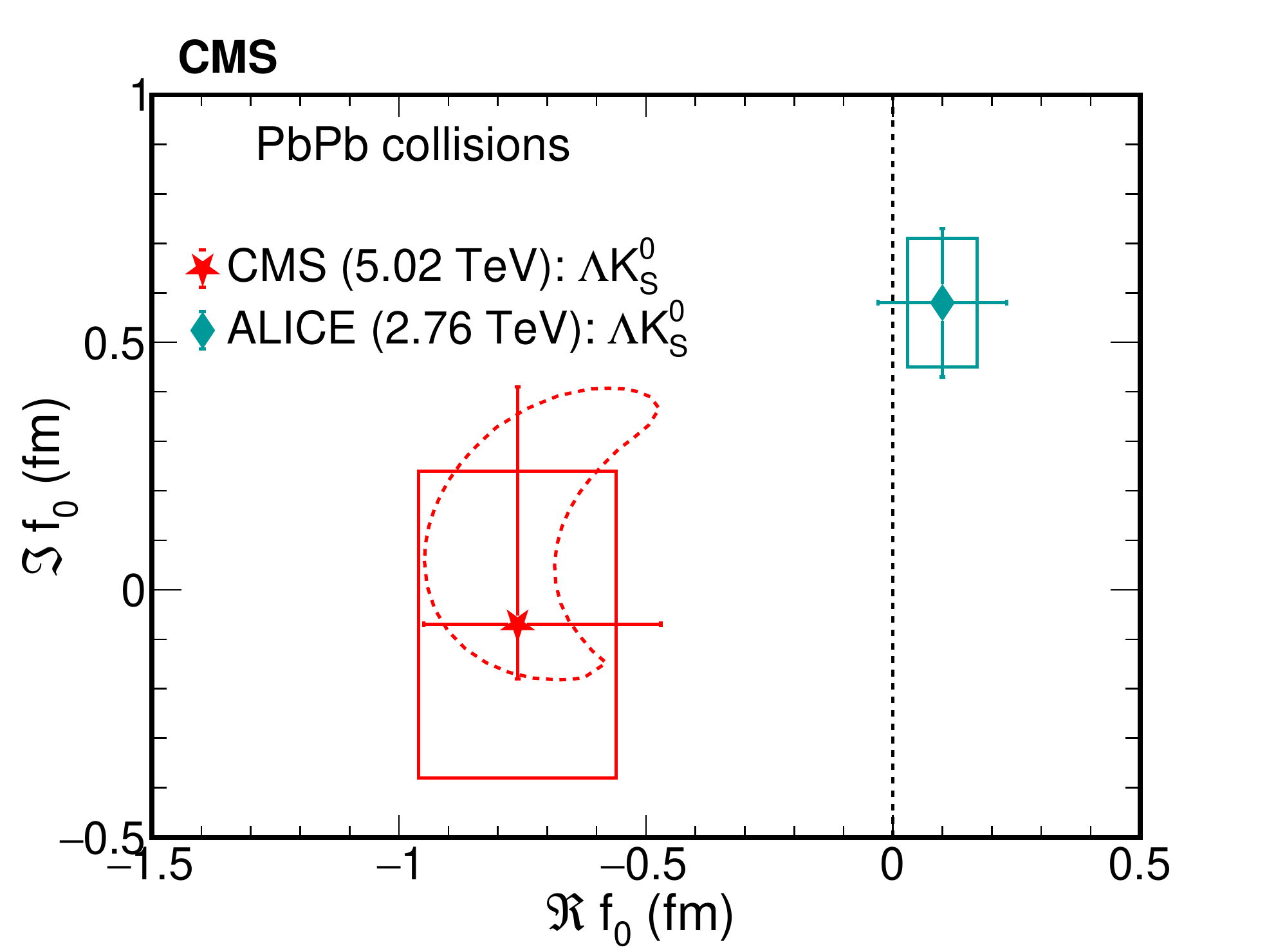}
\caption{The measured values of $d_0$ versus $\Re f_0$ (left) and $\Im f_0$ versus $\Re f_0$ (right) from this analysis along with other measurements and predictions as described in the text. For each data point, the lines and the boxes indicate the one-dimensional statistical and systematic uncertainties, respectively. For the CMS points, the dotted lines show two-dimensional contours corresponding to a 39\% confidence level with just the statistical uncertainties included.}
\label{fig:f0d0}
\end{figure*}

Table~\ref{tab:scatteringpara} also includes the strong interaction scattering parameters $d_0$, $\Re f_0$, and $\Im f_0$ obtained from the \LALKS and \LLALAL correlations. Figure~\ref{fig:f0d0} shows $d_0$ and $\Im f_0$ versus $\Re f_0$ in the left and right panels, respectively, along with comparisons to previous results. The current CMS results are shown as red stars and squares for \LALKS and \LLALAL, respectively. The boxes and lines indicate one-dimenssional statistical and systematic uncertainties, respectively. For the CMS results, a two-dimensional contour is also plotted to show where the $\chi^2$ value changes by one unit with respect to its minimum value. This corresponds to a 39\% confidence level, including only the statistical uncertainties.

\begin{table*}[ht]\renewcommand{\arraystretch}{1.2}
\topcaption{Extracted values of the \Rin, $\Re f_0$, $\Im f_0$, $d_0$, \Lam, and $\left<\mT\right>$ parameters from the \KSKS, \LALKS, and \LLALAL combinations in the 0--80\% centrality range. The first and second uncertainties are statistical and systematic, respectively.}
\centering
\begin{tabular}{ l  c  c  c }
Parameter  & \KSKS & \LALKS & \LLALAL \\
\hline
{\Rin (fm)} & $3.40\pm 0.11\pm 0.37$ & $2.1^{+1.4}_{-0.5}\pm 0.8$ & $1.3^{+0.4}_{-0.2}\pm 0.3$ \vspace{0.12cm}\\
{$\Re f_0$ (fm)} &  \NA & $-0.76^{+0.29}_{-0.19}\pm 0.20$ & $0.74^{+0.59}_{-0.16}\pm 0.33$ \vspace{0.12cm}\\
{$\Im f_0$ (fm)} &  \NA & $-0.07^{+0.48}_{-0.11}\pm 0.32$ & \NA \vspace{0.12cm}\\
{$d_0$ (fm)} & \NA & $2.3^{+0.7}_{-0.5}\pm 1.3$ & $4.2^{+5.7}_{-2.1}\pm 2.9$ \vspace{0.12cm}\\
{\Lam} & $0.43\pm 0.03\pm 0.13$ & $0.34^{+0.41}_{-0.12}\pm 0.17$ & $1.5^{+1.2}_{-1.1}\pm 1.4$ \vspace{0.12cm}\\
{$\left<\mT\right>$ ({\GeVns})} & 1.50 & 2.09 & 2.60 \vspace{0.12cm}\\
\end{tabular}
\label{tab:scatteringpara}
\end{table*}

The negative value of $\Re f_0$ observed for the \LALKS correlations, combined with its relatively small magnitude, suggests a repulsive \LKS interaction. The uncertainty associated with the $\Im f_0$ value for the \LALKS correlations prevents any claim concerning possible inelastic processes. The measured \LALKS values of $\Re f_0$, $\Im f_0$, and $d_0$ differ by about 2, 1, and 1 standard deviations, respectively, from the reported ALICE values (teal diamonds)~\cite{PhysRevC.103.055201}, which are obtained from PbPb collisions at $\sqrtsNN = 2.76\TeV$.  In contrast to this analysis, the ALICE result is corrected for the effect on the correlations caused by feed-down, which may be a source of the differences in the measured values.

The positive $\Re f_0$ value obtained for the \LLALAL correlations suggests an attractive interaction that is not strong enough to produce a bound state such as the H-dibaryon~\cite{PhysRevC.91.024916, STAR:2014dcy}. This result disagrees with the finding from the STAR Collaboration in gold-gold (\AuAu) collisions at $\sqrtsNN = 200\GeV$ (blue circle). The negative $\Re f_0$ value of ${-}1.10\pm 0.37\stat^{+0.68}_{-0.08}\syst$\unit{fm} found by STAR, combined with its magnitude, imply a repulsive interaction. It is noted, however, that a theoretical study of the STAR data which considers collective flow and feed-down effects (shown as a shaded region at $d_0\approx5$\unit{fm}, $\Re f_0\approx0.9$\unit{fm}) suggests that these data are consistent with the \LLALAL interaction being attractive~\cite{PhysRevC.91.024916}. An analysis by the ALICE Collaboration of the \LLALAL scattering parameters, using \LLALAL correlations from \pp collisions at $\roots = 7$ and 13\TeV, as well as \pPb collisions at $\sqrtsNN = 5.02\TeV$, excludes certain regions of the $d_0$ and $\Re f_0$ parameter space, while also suggesting an attractive interaction~\cite{ALICE:2019eol}. Our measured values fall within their allowed regions and provide further evidence for an attractive interaction. In addition, our results are consistent with two theoretical calculations (black triangles) that reproduce the \LLALAL binding energy of \He, as extracted from the NAGARA event~\cite{Hiyama_2002,Filikhin_2002}.

\section{Summary}
\label{sec:summary}
The \KSKS, \LALKS, and \LLALAL femtoscopic correlations are studied using lead-lead (\PbPb) collision data at a center-of-mass energy per nucleon pair of $\sqrtsNN = 5.02\TeV$, collected by the CMS Collaboration. This is the first report on \LLALAL correlations in \PbPb collisions at the CERN LHC. The source size \Rin and the correlation strength parameter \Lam were extracted for \KSKS correlations in six centrality bins covering the 0--60\% range. The value of \Rin decreases from 4.6 to 1.6\unit{fm} going from central to peripheral collisions and agrees with results from the ALICE Collaboration at a similar transverse mass. Along with the \Rin and \Lam parameters, the strong interaction scattering parameters, \ie, the complex scattering length and effective range, were extracted from \LALKS and \LLALAL correlations in the 0--80\% centrality range. These scattering parameters indicate that the \LALKS interaction is repulsive and that the \LLALAL interaction is attractive. The scattering parameters obtained from \LALKS correlations differ from those reported by the ALICE Collaboration. The positive real scattering length obtained from the \LLALAL correlation disfavors the existence of a bound H-dibaryon state. The \LLALAL scattering parameters help to constrain baryon-baryon and, more specifically, hyperon-hyperon interaction models. These measurements provide an additional input to understand the nature of the strong interaction between pairs of strange hadrons.

\begin{acknowledgments}
  We congratulate our colleagues in the CERN accelerator departments for the excellent performance of the LHC and thank the technical and administrative staffs at CERN and at other CMS institutes for their contributions to the success of the CMS effort. In addition, we gratefully acknowledge the computing centers and personnel of the Worldwide LHC Computing Grid and other centers for delivering so effectively the computing infrastructure essential to our analyses. Finally, we acknowledge the enduring support for the construction and operation of the LHC, the CMS detector, and the supporting computing infrastructure provided by the following funding agencies: BMBWF and FWF (Austria); FNRS and FWO (Belgium); CNPq, CAPES, FAPERJ, FAPERGS, and FAPESP (Brazil); MES and BNSF (Bulgaria); CERN; CAS, MoST, and NSFC (China); MINCIENCIAS (Colombia); MSES and CSF (Croatia); RIF (Cyprus); SENESCYT (Ecuador); MoER, ERC PUT and ERDF (Estonia); Academy of Finland, MEC, and HIP (Finland); CEA and CNRS/IN2P3 (France); BMBF, DFG, and HGF (Germany); GSRI (Greece); NKFIH (Hungary); DAE and DST (India); IPM (Iran); SFI (Ireland); INFN (Italy); MSIP and NRF (Republic of Korea); MES (Latvia); LAS (Lithuania); MOE and UM (Malaysia); BUAP, CINVESTAV, CONACYT, LNS, SEP, and UASLP-FAI (Mexico); MOS (Montenegro); MBIE (New Zealand); PAEC (Pakistan); MES and NSC (Poland); FCT (Portugal); MESTD (Serbia); MCIN/AEI and PCTI (Spain); MOSTR (Sri Lanka); Swiss Funding Agencies (Switzerland); MST (Taipei); MHESI and NSTDA (Thailand); TUBITAK and TENMAK (Turkey); NASU (Ukraine); STFC (United Kingdom); DOE and NSF (USA).
  
  \hyphenation{Rachada-pisek} Individuals have received support from the Marie-Curie program and the European Research Council and Horizon 2020 Grant, contract Nos.\ 675440, 724704, 752730, 758316, 765710, 824093, 884104, and COST Action CA16108 (European Union); the Leventis Foundation; the Alfred P.\ Sloan Foundation; the Alexander von Humboldt Foundation; the Belgian Federal Science Policy Office; the Fonds pour la Formation \`a la Recherche dans l'Industrie et dans l'Agriculture (FRIA-Belgium); the Agentschap voor Innovatie door Wetenschap en Technologie (IWT-Belgium); the F.R.S.-FNRS and FWO (Belgium) under the ``Excellence of Science -- EOS" -- be.h project n.\ 30820817; the Beijing Municipal Science \& Technology Commission, No. Z191100007219010; the Ministry of Education, Youth and Sports (MEYS) of the Czech Republic; the Hellenic Foundation for Research and Innovation (HFRI), Project Number 2288 (Greece); the Deutsche Forschungsgemeinschaft (DFG), under Germany's Excellence Strategy -- EXC 2121 ``Quantum Universe" -- 390833306, and under project number 400140256 - GRK2497; the Hungarian Academy of Sciences, the New National Excellence Program - \'UNKP, the NKFIH research grants K 124845, K 124850, K 128713, K 128786, K 129058, K 131991, K 133046, K 138136, K 143460, K 143477, 2020-2.2.1-ED-2021-00181, and TKP2021-NKTA-64 (Hungary); the Council of Science and Industrial Research, India; the Latvian Council of Science; the Ministry of Education and Science, project no. 2022/WK/14, and the National Science Center, contracts Opus 2021/41/B/ST2/01369 and 2021/43/B/ST2/01552 (Poland); the Funda\c{c}\~ao para a Ci\^encia e a Tecnologia, grant CEECIND/01334/2018 (Portugal); the National Priorities Research Program by Qatar National Research Fund; MCIN/AEI/10.13039/501100011033, ERDF ``a way of making Europe", and the Programa Estatal de Fomento de la Investigaci{\'o}n Cient{\'i}fica y T{\'e}cnica de Excelencia Mar\'{\i}a de Maeztu, grant MDM-2017-0765 and Programa Severo Ochoa del Principado de Asturias (Spain); the Chulalongkorn Academic into Its 2nd Century Project Advancement Project, and the National Science, Research and Innovation Fund via the Program Management Unit for Human Resources \& Institutional Development, Research and Innovation, grant B05F650021 (Thailand); the Kavli Foundation; the Nvidia Corporation; the SuperMicro Corporation; the Welch Foundation, contract C-1845; and the Weston Havens Foundation (USA).
  
  \end{acknowledgments}
\bibliography{auto_generated}
\cleardoublepage \appendix\section{The CMS Collaboration \label{app:collab}}\begin{sloppypar}\hyphenpenalty=5000\widowpenalty=500\clubpenalty=5000
\cmsinstitute{Yerevan Physics Institute, Yerevan, Armenia}
{\tolerance=6000
A.~Tumasyan\cmsAuthorMark{1}\cmsorcid{0009-0000-0684-6742}
\par}
\cmsinstitute{Institut f\"{u}r Hochenergiephysik, Vienna, Austria}
{\tolerance=6000
W.~Adam\cmsorcid{0000-0001-9099-4341}, J.W.~Andrejkovic, T.~Bergauer\cmsorcid{0000-0002-5786-0293}, S.~Chatterjee\cmsorcid{0000-0003-2660-0349}, K.~Damanakis\cmsorcid{0000-0001-5389-2872}, M.~Dragicevic\cmsorcid{0000-0003-1967-6783}, A.~Escalante~Del~Valle\cmsorcid{0000-0002-9702-6359}, P.S.~Hussain\cmsorcid{0000-0002-4825-5278}, M.~Jeitler\cmsAuthorMark{2}\cmsorcid{0000-0002-5141-9560}, N.~Krammer\cmsorcid{0000-0002-0548-0985}, L.~Lechner\cmsorcid{0000-0002-3065-1141}, D.~Liko\cmsorcid{0000-0002-3380-473X}, I.~Mikulec\cmsorcid{0000-0003-0385-2746}, P.~Paulitsch, J.~Schieck\cmsAuthorMark{2}\cmsorcid{0000-0002-1058-8093}, R.~Sch\"{o}fbeck\cmsorcid{0000-0002-2332-8784}, D.~Schwarz\cmsorcid{0000-0002-3821-7331}, M.~Sonawane\cmsorcid{0000-0003-0510-7010}, S.~Templ\cmsorcid{0000-0003-3137-5692}, W.~Waltenberger\cmsorcid{0000-0002-6215-7228}, C.-E.~Wulz\cmsAuthorMark{2}\cmsorcid{0000-0001-9226-5812}
% ...
\par}
\vskip\cmsinstskip
\dag:~Deceased\\
$^{1}$Also at Yerevan State University, Yerevan, Armenia\\
$^{2}$Also at TU Wien, Vienna, Austria\\
% ...
\end{sloppypar}
\end{document}